\documentstyle[12pt]{article}
\def\ovl{\overline}
\def\bra{\langle}
\def\ket{\rangle}

\newcommand{\eqnapp}{\setcounter{equation}{0}%
      \renewcommand{\theequation}{A.\arabic{equation}}}%

\begin{document}
\vspace{.4cm}
\begin{flushright}
\large{HUTP-97/A011\\
March 1997\\
version 0.1}
\end{flushright}
\vspace{0.5cm}
\begin{center}
\LARGE{\bf Lepton $CP$ Asymmetries in B Decays}
\end{center}

\vspace{1.2cm}
\begin{center}\large
Hitoshi Yamamoto\\
\vspace{.4cm}
{\normalsize 
{\it Dept. of Physics, Harvard University, 42 Oxford St., Cambridge, 
MA 02138, U.S.A.}\\
{\it e-mail: yamamoto@huhepl.harvard.edu}}
\vspace{.4cm}
\end{center}

\thispagestyle{empty}
\vspace{3.4cm}
\centerline{\Large \bf Abstract}
\vspace{0.3cm}
In the decay of $\Upsilon(4S)$, the decay time distribution of
$\Upsilon(4S)\to f + X$, where $f$ is a final state 
that $B^0$ or $\ovl B^0$
can decay to, is the sum of the decay time distributions of
$B^0\to f$ and $\ovl B^0 \to f$. Using this general rule, we estimate the
sensitivity of single lepton $CP$ violation measurements with respect to that
of traditional di-lepton measurements. We find that the
sensitivity of the single lepton method 
is comparable to or better than that of the
di-lepton method. The two data samples are largely statistically independent,
so that they can be combined to improve sensitivity.
The advantage of the single lepton measurement
increases for large mixings, which suggests that the single lepton
method holds promise for $B_s$.
We also discuss lepton asymmetry measurements
on the $Z^0$ peak and in hadron colliders.
The achievable
sensitivity with the currently available data is already in the range 
relevant to standard
model predictions. 

\vfill

\newpage

\addtocounter{page}{-1}

\section{Introduction}

One of the ways $CP$ violation can manifest itself is in the 
particle-antiparticle imbalance in mass eigenstates of
neutral meson systems.
In the case of the neutral kaon, it can be measured as the asymmetry between
$K_L\to \pi^- \ell^+ \nu$ and $K_L\to \pi^+ \ell^- \nu$ \cite{PDG,Kaon}:
\begin{equation}
    \delta_K\equiv
      {Br(K_L\to \pi^- \ell^+ \nu) - Br(K_L\to \pi^+ \ell^- \nu) \over
       Br(K_L\to \pi^- \ell^+ \nu) + Br(K_L\to \pi^+ \ell^- \nu)}
     = (3.27\pm0.12)\times10^{-3}\;,
\end{equation}
which indicates that $K_L$ contains more $K^0$ than $\ovl K^0$ 
(assuming $\Delta S = \Delta Q$). If
$CPT$ is conserved, $K_S$ has the same asymmetry $\delta_K$
with the same sign; thus there is no need to specify which of the
two mass eigenstates is being considered. 

For the neutral $B$ meson
system, one can similarly define the asymmetry $\delta$ as
\begin{equation}
    \delta\equiv     
      {|\bra B^0|B_{a,b}\ket|^2 - |\bra\ovl B^0|B_{a,b}\ket|^2\over
       |\bra B^0|B_{a,b}\ket|^2 + |\bra\ovl B^0|B_{a,b}\ket|^2}\; ,
   \label{eq:deltadef}
\end{equation}
where $B_a$ and $B_b$ are the two mass eigenstates which, as in the case
of the kaon, have the same asymmetry (assuming $CPT$). In practice, however,
it is experimentally difficult to isolate $B_a$ or $B_b$. The traditional
method is to measure the same-sign di-lepton asymmetry in 
$\Upsilon(4S)\to B^0 \ovl B^0$ \cite{Okun+,AliAydin}:
\begin{equation}
     A_{\ell\ell}\equiv
    { N(\ell^+\ell^+) - N(\ell^-\ell^-) \over
      N(\ell^+\ell^+) + N(\ell^-\ell^-) } 
      \sim 2\delta\; .
\end{equation}
There is in principle no dilution due to $\Upsilon(4S)\to B^+B^-$ since
the same-sign di-lepton events are caused by mixing of the neutral $B$ meson
(assuming $\Delta B = \Delta Q$).

Within the framework of the standard model, the short distance calculation
gives \cite{Ma+,Barger+,Hagelin}
\begin{equation}
     A_{\ell\ell}\sim -\,4\pi{m_c^2\over m_t^2}\,
       \Im\left({V_{cb}V_{cd}^*\over V_{tb}V_{td}^*}\right),
    \label{Ashort}
\end{equation}
which is around $10^{-3}$. Including long distance effects, Altomari {\it et al.}
estimated $A_{\ell\ell}$ to be in the range $10^{-3}$ to $10^{-2}$
\cite{Altomari+}.  The uncertainty is primarily due to
hadronic intermediate states,
and even the sign cannot be reliably predicted. 
As a consequence, a measurement of $CP$ violation in the semileptonic
asymmetry does not lead to the determination of basic $CP$ violating
parameters in the standard model. 
Outside of the standard model, however, the asymmetry can
be larger, and thus an experimental value of 
$|\delta|$ above $\sim10^{-2}$ would signal 
new physics \cite{Asl-Newphys,Newphys}. The current experimental number is
not very recent or precise: 
$A_{\ell\ell} = 0.031\pm0.096\pm0.032$ \cite{CLEO-A}.
For $B_s$, the short distance prediction of $A_{\ell\ell}$ is obtained by
replacing $d$ by $s$ in (\ref{Ashort}), and it is
even smaller than for $B^0$.

The $CP$ asymmetry in single lepton sample 
had been suggested as a possible observable to search for $CP$ violation
in the case when the mixing is small \cite{Hagelin,Buras+}.
The logic was that if the mixing is
small, then the statistics of the di-lepton events will decrease,
making the di-lepton method impractical.
After the observation of substantial mixing in the neutral 
$B$ meson system \cite{Bmixobs}, however, the single lepton method 
has not received much attention. 
In this note, we point out that the advantage of the single lepton method
over the di-lepton method actually increases for large mixings, and that,
on $\Upsilon(4S)$, the single lepton method has a comparable or 
better sensitivity than the di-lepton method. This is so in spite of the fact
that in the single lepton measurement, one usually cannot distinguish charged
and neutral $B$ mesons.
We begin by briefly reviewing the phenomenological
background, and then move on to estimate experimental sensitivities.
In the appendix, we present a general rule that relates the inclusive
decay time distribution on $\Upsilon(4S)$ to those of $B^0$ and $\ovl B^0$
tagged at $t=0$, as well as decay rate formulas without assuming $CPT$
invariance.

\section{Phenomenology}

The mass
eigenstates can be written in terms of $B^0$ and $\ovl B^0$ as
\begin{equation}
   \left\{
   \begin{array}{rcl@{\quad}l}
      B_a &=& p  B^0 + q  \ovl B^0 
            &(\hbox{mass: }m_a,\hbox{ decay rate: }\gamma_a)\\
      B_b &=& p' B^0 - q' \ovl B^0 
            &(\hbox{mass: }m_b,\hbox{ decay rate: }\gamma_b)\\
   \end{array}, \right.
   \label{eq:Babdef}
\end{equation}
where the normalization is
\begin{equation}
   |p|^2 + |q|^2 = 1\; ,\qquad |p'|^2 + |q'|^2 = 1\; .
\end{equation}
If $CPT$ is a good symmetry, we have
\begin{equation}
   p' = p\; , \quad q' = q \quad (CPT)\; . \label{eq:CPTpq}
\end{equation}
In the following we assume $CPT$ invariance. $CPT$ symmetry
also effectively allows us to take \cite{CPTamp}
\begin{equation}
   |Amp(B^0\to\ell^+)| = |Amp(\ovl B^0\to\ell^-)|\equiv A_0\quad (CPT)\; .
        \label{eq:CPTamp}
\end{equation}
Furthermore, we will assume $\Delta B = \Delta Q$ \cite{dBdQ}. 
The probability that a pure $B^0$ 
or $\ovl B^0$ at $t=0$ decays to $\ell^\pm$ at time $t$ is then given by
(see Appendix)
\begin{eqnarray}
  \Gamma_{B^0\to\ell^+}(t) &=& \Gamma_{\ovl B^0\to\ell^-}(t) =
  {A_0^2\over2} e^{-\gamma_+ t} 
       \left[\cosh\gamma_- t + \cos\delta m t\right]\; , \nonumber \\
  \Gamma_{\ovl B^0\to\ell^+}(t) &=& {|p|^2\over |q|^2}
  {A_0^2\over2} e^{-\gamma_+ t}
       \left[\cosh\gamma_- t - \cos\delta m t\right]\; , \label{eq:BBlepdist}\\
  \Gamma_{ B^0\to\ell^-}(t)     &=& {|q|^2\over |p|^2}
  {A_0^2\over2} e^{-\gamma_+ t}
       \left[\cosh\gamma_- t - \cos\delta m t\right]\; , \nonumber
\end{eqnarray}
where
\begin{equation}
    \delta m\equiv m_a - m_b\; ,\quad
    \gamma_\pm \equiv {\gamma_a\pm\gamma_b\over 2}\; .
\end{equation}
The short distance calculation predicts
$\gamma_a\sim\gamma_b$ \cite{Buras+}.
There is, however, no stringent experimental limit, and we will allow for the
possibility that the difference is sizable. Also, the following expressions
are applicable to $B_s$ mesons which is expected to have a sizable
decay rate difference.
The fraction of $B^0$ 
or $\ovl B^0$ at $t=0$ eventually decaying to $\ell^\pm$, which we denote
as $Br(B^0 (\ovl B^0) \to\ell^\pm)$, is obtained by integrating
the above expressions:
\begin{eqnarray}
    Br(B^0\to\ell^+)      &=& Br(\ovl B^0\to\ell^-) = 
          {b_{sl}\over 1-y^2}\, (1-\chi)\; ,  \nonumber \\
    Br(\ovl B^0\to\ell^+) &=& 
     {b_{sl}\over 1-y^2}{|p|^2\over |q|^2}\,\chi\; , \label{eq:BBlep}\\
    Br(B^0\to\ell^-)      &=& 
     {b_{sl}\over 1-y^2}{|q|^2\over |p|^2}\,\chi\; ,   \nonumber 
\end{eqnarray}
where $\chi$ is the standard mixing parameter \cite{PaisTreiman} 
defined by
\begin{equation}
  \chi \equiv {1\over2}\,{x^2 + y^2\over 1 + x^2}
    = 0.175\pm0.016\;\cite{PDG}\;,
\end{equation}
\begin{equation}
    x \equiv {\delta m\over\gamma_+}\;,\quad 
    y \equiv {\gamma_-\over\gamma_+}\;,
\end{equation}
and $b_{sl}$ is the `normalized' semileptonic branching fraction
which reduces to the experimental semileptonic branching fraction
in the limit of $\gamma_a = \gamma_b$ and $CP$ symmetry:
\begin{equation}   
     b_{sl}\equiv {\displaystyle {A_0^2\over\gamma_+}} \;
     \stackrel{\gamma_a=\gamma_b, CP}{\longrightarrow} \;
     2 Br(B\to X\ell\nu) = 2 \times (0.1043\pm0.0024)\;\cite{PDG}\;,
   \label{eq:SLamp}
\end{equation}
where the factor 2 comes from the fact that there are electron and muon modes.
From the first line of (\ref{eq:BBlep}), we see that there is no asymmetry 
in the `right-sign' lepton branching
fractions. 
The particle-antiparticle imbalance in $B_{a,b}$ (namely,
$|p|^2 \not= |q|^2$) shows up only in the `wrong-sign' decays:
\begin{equation}
   {Br(\ovl B^0\to\ell^+) - Br(B^0\to\ell^-) \over
    Br(\ovl B^0\to\ell^+) + Br(B^0\to\ell^-)}
    = {|p|^4 - |q|^4\over |p|^4 + |q|^4} = {2\delta \over 1 + \delta^2}\;,
   \label{eq:Aflvtag}
\end{equation}
with
\begin{equation}
     \delta = {|p|^2 - |q|^2\over |p|^2 + |q|^2}\; ,
\end{equation}
which follows from definition (\ref{eq:deltadef}).
Experimentally, measurement of this asymmetry requires flavor tagging
at $t=0$. 

In the decay $\Upsilon(4S)\to B^0\ovl B^0$, one needs to take into account
the quantum correlations arising from the fact that the two mesons are in
a coherent $L=1$ state. 
We consider the case when one side decays to $\ell^+$ and the other side
decays to $\ell^-$ (and all other charge combinations).
The decay time variable accessible in an asymmetric $B$-factory is the
time difference of decays:
\begin{equation}
        t_- \equiv t_1 - t_2\; .
\end{equation}
In terms of $t_-$, the decay time distributions are (see Appendix)
\begin{eqnarray}
    \Gamma_{\Upsilon(4S)\to\ell^+\ell^-}(t_-) &=& 
    \Gamma_{\Upsilon(4S)\to\ell^-\ell^+}(t_-) \nonumber  \\
     &=& {A_0^4\over8\gamma_+}  e^{-\gamma_+ |t_-|}
       \left[ \cosh\gamma_- t_- + \cos \delta m t_- \right] \;, 
                                               \nonumber     \\
    \Gamma_{\Upsilon(4S)\to\ell^+\ell^+}(t_-) &=& 
         {|p|^2\over|q|^2}{A_0^4\over8\gamma_+}  e^{-\gamma_+ |t_-|}
       \left[ \cosh\gamma_- t_- - \cos \delta m t_- \right]\;,
                                          \label{eq:Upstmi}\\ 
    \Gamma_{\Upsilon(4S)\to\ell^-\ell^-}(t_-) &=& 
         {|q|^2\over|p|^2}{A_0^4\over8\gamma_+}  e^{-\gamma_+ |t_-|}
       \left[ \cosh\gamma_- t_- - \cos \delta m t_- \right]\; .
                                        \nonumber
\end{eqnarray}
Note that  the $\ell^+\ell^+$ and $\ell^-\ell^-$ distributions 
have exactly the
same $t_-$ dependence, and thus the asymmetry between them does not
depend on $t_-$.
We thus integrate (\ref{eq:Upstmi}) to obtain
the fraction of
$\Upsilon(4S)\to B^0\ovl B^0$ decays that eventually result in 
a lepton pair $\ell^\pm\ell^\pm$, which we denote by 
$Br(\Upsilon(4S)\to \ell^\pm\ell^\pm)$:
\begin{eqnarray}
    Br(\Upsilon(4S)\to\ell^+\ell^-)  &=&  Br(\Upsilon(4S)\to\ell^-\ell^+)
        = {1\over2}{b_{sl}^2\over 1-y^2}\, (1-\chi)\;,  \nonumber \\
    Br(\Upsilon(4S)\to\ell^+\ell^+) &=& 
          {1\over2}{b_{sl}^2\over 1-y^2}{|p|^2\over |q|^2}\,\chi\;, 
                                                  \label{eq:Upslep}\\
    Br(\Upsilon(4S)\to\ell^-\ell^-)      &=& 
          {1\over2}{b_{sl}^2\over 1-y^2}{|q|^2\over |p|^2}\,\chi\;.
                                                  \nonumber 
\end{eqnarray}
The common factor $b_{sl}^2 / 1-y^2$ can be written as
\begin{equation}
    {b_{sl}^2\over 1-y^2} =
    Br(B_a\to\ell^\pm)\, Br(B_b\to\ell^\pm)\;.
\end{equation} 
Then, in the absence of $CP$ violation (namely, $|p|^2=|q|^2$), 
the total yield of di-lepton events
becomes $Br(B_a\to\ell^\pm) Br(B_b\to\ell^\pm)$, as expected from the
simple picture $\Upsilon(4S)\to B_a B_b$. 
The asymmetry between $(\ell^+\ell^+)$ and $(\ell^-\ell^-)$ is the same
as that of the `wrong-sign' leptons discussed above:
\begin{equation}
    A_{\ell\ell} \equiv
   {Br(\Upsilon(4S)\to\ell^+\ell^+) - Br(\Upsilon(4S)\to\ell^-\ell^-) \over
    Br(\Upsilon(4S)\to\ell^+\ell^+) + Br(\Upsilon(4S)\to\ell^-\ell^-) }
    = {2\delta \over 1 + \delta^2}\;.
    \label{eq:Asymll}
\end{equation}
No flavor tagging is required for this measurement.

In order to study the single lepton asymmetry on $\Upsilon(4S)$, one 
has to take into account the cases where one side decays semileptonically
and the other side decays to anything. To do so, we use the
following general rule (see Appendix):
\begin{equation}
    \Gamma_{\Upsilon(4S)\to f}(t) =
    2 \sum_{f_1}\int_0^\infty 
      \Gamma_{\Upsilon(4S)\to f_1f}(t_1,t)\, dt_1 = 
      \Gamma_{B^0\to f}(t) + \Gamma_{\ovl B^0\to f}(t)\;,
    \label{eq:4Sinc}
\end{equation}
where $\Gamma_{\Upsilon(4S)\to f}(t)$ is the probability density
that one finds a given final state $f$ decaying at time $t$ in the
process $\Upsilon(4S)\to B^0\ovl B^0$, 
$\Gamma_{\Upsilon(4S)\to f_1f_2}(t_1,t_2)$ 
is the probability density that
one side of $\Upsilon(4S)\to B^0\ovl B^0$ 
decays to final state $f_1$ at time $t_1$ 
and the other side decays
to $f_2$ at time $t_2$, and $\Gamma_{B^0(\ovl B^0)\to f}(t)$
is the probability density that a pure $B^0(\ovl B^0)$ at $t=0$ 
decays to final state $f$ at time $t$. The factor 2 in 
(\ref{eq:4Sinc}) accounts for the fact that the final state $f$
can come from either side of the $\Upsilon(4S)$ decay. 
These functions are
related to the branching fractions discussed above by
\begin{equation}
    Br(\Upsilon(4S)\to f_1f_2) =
     \int_0^\infty \int_0^\infty
     \Gamma_{\Upsilon(4S)\to f_1f_2}(t_1,t_2)\, dt_1dt_2\; ,
\end{equation}
\begin{equation}
    Br(B^0(\ovl B^0)\to f) =
     \int_0^\infty \Gamma_{B^0(\ovl B^0)\to f}(t)\,dt\;.
   \label{eq:BBgamma}
\end{equation}
The relation (\ref{eq:4Sinc}) is a consequence of quantum mechanics
and conservation of probability, and is valid even when $CPT$ is violated.
The same relation holds for $e^+e^-\to V \to K^0\ovl K^0$,
$D^0\ovl D^0$, and $B_s \ovl B_s$, where $V$ is a vector state or
a virtual photon.

Let us define the inclusive quantity on $\Upsilon(4S)$ by
\begin{equation}
    N(\Upsilon(4S)\to f) \equiv 
     \int_0^\infty\Gamma_{\Upsilon(4S)\to f}(t)\, dt\; ,
  \label{eq:4SBr}
\end{equation}
which is the total expected number of final state $f$ in one
$\Upsilon(4S)\to B^0\ovl B^0$ decay.
The normalizations are given by (see Appendix)
\begin{eqnarray}
 \sum_{f_1,f_2} Br(\Upsilon(4S)\to f_1f_2) &=& 1\;, \label{eq:Upsnorm}\\
 \sum_f Br(B^0(\ovl B^0)\to f) &=& 1\;, \label{eq:BBnorm} \\
 \sum_f N(\Upsilon(4S)\to f) &=& 2 \;. \label{eq:Ups1norm}
\end{eqnarray}
The last normalization reflects the fact that there are two
$B$ mesons per $\Upsilon(4S)$ decay.

From (\ref{eq:BBlep}),(\ref{eq:4Sinc}),(\ref{eq:BBgamma}), and
(\ref{eq:4SBr}), one then obtains the inclusive lepton yields on
$\Upsilon(4S)$:
\begin{eqnarray}
   N(\Upsilon(4S)\to \ell^+) &=& {b_{sl}\over 1-y^2}
       \left[ 1 + \left( {|p|^2\over|q|^2}-1\right)\chi\right] \;,
                    \nonumber  \\
   N(\Upsilon(4S)\to \ell^-) &=& {b_{sl}\over 1-y^2}
       \left[ 1 + \left( {|q|^2\over|p|^2}-1\right)\chi\right]\; .
             \label{eq:Ups1lep}
\end{eqnarray}
One sees that when $|p|^2\not= |q|^2$ there is an asymmetry. In practice,
however, leptons from $B^\pm$ are difficult to reject, and the resulting
dilution needs to be taken into account.

\section{Experimental Sensitivities}

We will now estimate the sensitivities to $\delta$ of single and
di-lepton asymmetry measurements. We assume that the lepton detection
efficiency $\epsilon_\ell$ for each lepton 
is the same in the single and di-lepton cases,
and that they are uncorrelated in the latter.
Also we assume $\delta\ll 1$ for the expressions of
asymmetries below. In estimating statistics, we further assume
$\gamma_a\sim\gamma_b$ (or equivalently 
$y\ll 1$).
If we have $N_0$ $\Upsilon(4S)\to B^0\ovl B^0$ decays, then from 
(\ref{eq:Upslep}) the total number of same sign di-lepton events
detected is $N_0\, b_{sl}^2\,\chi\,\epsilon_\ell^2$. 
Using (\ref{eq:Asymll}), the error in
$\delta$ is then
\begin{equation}
   \sigma_{\delta}(\ell\ell) = {1\over2}
         {1\over\sqrt{N_0\, b_{sl}^2\,\chi\,\epsilon_\ell^2}}\;.
\end{equation}
The single lepton asymmetry on $\Upsilon(4S)$ 
can be obtained from (\ref{eq:Ups1lep}):
\begin{equation}
    A_\ell(\Upsilon(4S)) \equiv D\;
     {N(\Upsilon(4S)\to\ell^+) - N(\Upsilon(4S)\to\ell^-) \over
      N(\Upsilon(4S)\to\ell^+) + N(\Upsilon(4S)\to\ell^-) } 
    = 2 D\,\chi\,\delta\;,
\end{equation}
where $D$ is the dilution factor due to charged
$B$ mesons, and is equal to the fraction of leptons coming from neutral $B$
mesons. Other dilution effects such as those due to misidentified
leptons or leptons from charmed hadrons could also be absorbed into $D$. 
Assuming that there are the same number of leptons from
charged $B$'s as from neutral $B$'s, we take $D=1/2$.
The total number of single lepton events detected is
$N_0\, 4 b_{sl}\,\epsilon_\ell$; thus the sensitivity to $\delta$
of the single lepton measurement is
\begin{equation}
    \sigma_{\delta}(\ell) = {1\over\chi}
       {1\over\sqrt{N_0\, 4 b_{sl}\,\epsilon_\ell}}\;.
\end{equation}
The ratio of sensitivities of single to di-lepton measurements is then
\begin{equation}
  {\sigma_{\delta}(\ell)\over\sigma_{\delta}(\ell\ell)} =
    \sqrt{{b_{sl}\,\epsilon_\ell\over\chi}}\; .
\end{equation}
We see that the larger the mixing, the more advantageous the single
lepton method becomes. This may be counter-intuitive, but can be
understood as follows: as $\chi$ increases, the
statistics goes up linearly for
the di-lepton sample while its asymmetry stays the same.  
For the single lepton sample, 
the statistics stays the same while the asymmetry goes
up linearly, which is equivalent to statistics increasing quadratically for
a fixed asymmetry.

A typical value for $\epsilon_\ell$ is 0.5. Together with the experimental
values for $b_{sl}$ and $\chi$, the ratio above is 0.78. Namely, the single
lepton measurement has a sensitivity comparable to or better than 
that of the di-lepton
measurement. Note also that the
single and di-lepton datasets are largely statistically independent
(only about 10\%\ of the single lepton events are also in the di-lepton
dataset). The two measurements can thus be combined to improve
overall sensitivity. For example, the current CLEO data corresponds
to $N_0\sim 2\times10^6$. This
gives $\sigma_{\delta}(\ell) = 0.6\%$ and 
$\sigma_{\delta}(\ell\ell) = 0.8\%$ with the combined sensitivity of
0.5\% which is already in the range relevant to 
standard model predictions.
 
When $B^0$'s and $\ovl B^0$'s are generated incoherently (e.g. on
$Z^0$ or in $p\bar p$ collisions), one cannot perform the correlated
di-lepton analysis. However, one
can still perform single lepton asymmetry measurements.
The discussion below applies also to $B_s$ mesons.
When an equal number of $B^0$ and $\ovl B^0$ are generated, the decay
time distribution of $\ell^+$ can be compared to that of $\ell^-$
without tagging the flavor of the parent $B$ meson. Using
(\ref{eq:BBlepdist}), 
the time dependent single lepton asymmetry is then
\begin{equation}
  A_\ell(t)\equiv D\;{\Gamma_{B^0,\ovl B^0\to\ell^+}(t) -
                  \Gamma_{B^0,\ovl B^0\to\ell^-}(t) \over
                  \Gamma_{B^0,\ovl B^0\to\ell^+}(t) +
                  \Gamma_{B^0,\ovl B^0\to\ell^-}(t)} =
     D\, \delta \left(1 - {\cos\delta mt \over\cosh\gamma_- t}\right)\;,
   \label{eq:ABBleptim}
\end{equation}
where $D$ is the dilution factor due to $B^\pm$'s as before.
In this case, $D$ could in principle
be a function of decay time (e.g., if the lifetimes of neutral and
charged $B$ mesons are different). Because of the relation (\ref{eq:4Sinc}),
the time dependent asymmetry for the single lepton measurement on
$\Upsilon(4S)$ is also given by (\ref{eq:ABBleptim}).
We see that the asymmetry starts out as zero at $t=0$ and reaches the
first maximum at around $\delta m t\sim\pi$ (about 4 times the $b$
lifetime). If we simply count the number of leptons without 
measuring the decay time, then the asymmetry 
becomes the same as that on $\Upsilon(4S)$:
\begin{equation}
  A_\ell \equiv D\;
   {Br(B^0,\ovl B^0\to\ell^+) - Br(B^0,\ovl B^0\to\ell^-) \over
    Br(B^0,\ovl B^0\to\ell^+) + Br(B^0,\ovl B^0\to\ell^-)} =
  2 D\, \chi\,\delta\; .
\end{equation}
where we have used (\ref{eq:BBlep}). The factor $D$ again includes
the dilution due to $B^\pm$'s.
This observable (with $D=1$) was first proposed by Hagelin \cite{Hagelin} as a
measure of observability of $CP$ violation in $B^0-\ovl B^0$ mixing, since it
contains both the mixing parameter $\chi$ and the $CP$ violation parameter
$\delta$. 

The currently available statistics on $Z^0$ is 
smaller than that of CLEO; the $p\bar p$
collider at Fermilab, however, may be able to obtain a better sensitivity.
In actual data analysis,
the decay time is often required to 
be larger than a given threshold in order to
reject non-$B$ background. Such a requirement, however, should not
sacrifice sensitivity significantly since the asymmetry
at short decay time is small as seen in (\ref{eq:ABBleptim}).
Also, there is a possibility that vertexing allows 
separation of neutral $B$'s from charged $B$'s by counting the total
charge emerging from a given vertex, which would substantially
improve sensitivity. 
In addition, flavor tagging by leptons, jet charge, 
or associated pion production
\cite{CDFtag} may allow for measurement of the flavor-tagged
asymmetry (\ref{eq:Aflvtag}).
The vertexing technique may become useful also at
asymmetric $B$-factories.

In conclusion, we have studied the sensitivity of single lepton $CP$
asymmetry relative to that of the traditional di-lepton asymmetry
on $\Upsilon(4S)$. We find that the single lepton sensitivity is 
comparable to or better than that of the di-lepton analysis.
The achievable sensitivities on $\Upsilon(4S)$ and in $p\bar p$ collisions
with currently available datasets
are already close to the predictions of the standard model. 
The single lepton method also holds promise for 
measurement of the leptonic $CP$ asymmetry of $B_s$. 
In the near future (namely, at the $B$-factories, HERA-$B$, and
the upgraded $p\bar p$ collider), it is quite
possible that 
$CP$ violation will be observed in the semileptonic modes.

\vspace{1cm}
\noindent {\Large\bf Acknowledgements}
\vspace{0.5cm}

\noindent I would like to thank Sheldon Glashow for pointing out the
asymmetries in the single lepton counting, and
Isi Dunietz for useful discussions.
This work was supported by the Department of Energy
Grant DE-FG02-91ER40654.

\vspace{1cm}
\noindent {\Large\bf Appendix} \eqnapp%
\vspace{0.5cm}

\noindent This section is based on
quantum mechanics, conservation of probability, and the 
Weisskopf-Wigner formalism \cite{Wigner+}. We will not assume
$CPT$ invariance unless otherwise stated. No further approximations
are made. 
Solving (\ref{eq:Babdef}) for $B^0$ and $\ovl B^0$, we obtain
\begin{equation}
   \left\{
   \begin{array}{rcl}
          B^0 &=& c\, ( q' B_a + q B_b ) \\ 
     \ovl B^0 &=& c\, ( p' B_a - p B_b ) \\ 
   \end{array}
   \right. ,
\end{equation}
with
\begin{equation}
    c \equiv {1\over p'q + pq'}\;.
\end{equation}
The time evolution of the mass eigenstates are given by
\begin{equation}
          B_a \to e^{-(\gamma_a/2 + i m_a)t} B_a\; , \qquad
          B_b \to e^{-(\gamma_b/2 + i m_b)t} B_b\; .
\end{equation}
If we have a pure $B^0$ or $\ovl B^0$ at $t=0$, 
the decay time distributions to a final state $f$ are then
\begin{eqnarray}
   \Gamma_{B^0\to f}(t) &=& |c|^2 
        \left[ |q'a_f|^2 e^{-\gamma_a t} +
               |q b_f|^2 e^{-\gamma_b t} +
             2 \Re((q'a_f)^*(q b_f) e^{-(\gamma_+ - i\delta m)t}) \right] ,
           \nonumber \\
   \Gamma_{\ovl B^0\to f}(t) &=& |c|^2 
        \left[ |p'a_f|^2 e^{-\gamma_a t} +
               |p b_f|^2 e^{-\gamma_b t} -
             2 \Re((p'a_f)^*(p b_f) e^{-(\gamma_+ - i\delta m)t}) \right] ,
     \label{eq:BBdist}
\end{eqnarray}
where
\begin{equation}
    a_f \equiv Amp(B_a\to f)\;,\quad  b_f \equiv Amp(B_b\to f)\;.
\end{equation}
The parameters $\gamma_\pm$ and $\delta m$ are defined in the main text.
Note that we would have $|c|^2 = 1$ if $CPT$ and $CP$ were conserved.
The normalization of the decay amplitudes is such that
\begin{equation}
  \sum_f {|a_f|^2\over\gamma_a} = 1\;, \quad
  \sum_f {|b_f|^2\over\gamma_b} = 1\; .
    \label{eq:ampnorm}
\end{equation}
Integrating (\ref{eq:BBdist}) over time gives the 
fraction of a pure $B^0$ or $\ovl B^0$
at $t=0$ eventually decaying to a final state $f$:
\begin{eqnarray}
  Br(B^0\to f) &=& |c|^2\left[ |q'|^2{|a_f|^2\over\gamma_a} +
                               |q |^2{|b_f|^2\over\gamma_b} +
   2\Re\left(q'^*q{a_f^*b_f\,\over\gamma_+ - i\delta m}\right)\right] ,
      \nonumber \\
  Br(\ovl B^0\to f) &=& |c|^2\left[ |p'|^2{|a_f|^2\over\gamma_a} +
                                    |p |^2{|b_f|^2\over\gamma_b} -
   2\Re\left(p'^*p{a_f^*b_f\,\over\gamma_+ - i\delta m}\right)\right] .
\end{eqnarray}
The normalization (\ref{eq:BBnorm})
can be obtained by summing the above equations over $f$ and using 
the Bell-Steinberger 
relation\cite{Bell-Stein}
\begin{equation}
     {\sum_f a_f^*b_f\over\gamma_+ - i\delta m} =
     \bra B_a | B_b \ket\;\; ( =  p'p^* - q'q^* )\; ,
   \label{eq:Bell-Stein}
\end{equation}
which expresses the conservation of probability.

On $\Upsilon(4S)$, the $B^0\ovl B^0$ pair is created in the coherent
$L=1$ state
\begin{equation}
  \begin{array}{l}
    {\displaystyle {1\over\sqrt2}}
      \left( |\ovl B^0 (1)\ket |B^0 (2)\ket -
                  |B^0 (1)\ket |\ovl B^0 (2)\ket \right) \\
  \qquad\qquad = {\displaystyle {c\over\sqrt2}}
      \left( |B_a (1)\ket |B_b (2)\ket -
             |B_b (1)\ket |B_a (2)\ket \right)\; ,
  \end{array}
\end{equation}
where the numbers 1 and 2 distinguish the sides; namely, they
may be distinguished by the direction of the $B$ meson in the $\Upsilon(4S)$
C.M. system: $\hat k$ or $-\hat k$. Then the probability density
that the side 1 decays to final state $f_1$ at time $t_1$ and 
the side 2 decays to $f_2$ at time $t_2$ is given by
\begin{eqnarray}
   \Gamma_{\Upsilon(4S)\to f_1f_2}(t_1,t_2) &=& {|c|^2\over2}
\left[ e^{-\gamma_a t_1 -\gamma_b t_2} |a_{f_1}b_{f_2}|^2 +
       e^{-\gamma_b t_1 -\gamma_a t_2} |b_{f_1}a_{f_2}|^2 
    \right.\nonumber\\ 
   &&\left. \; -
   2 \Re\left( e^{-(\gamma_+ - i \delta m) t_1} 
               e^{-(\gamma_+ + i \delta m) t_2} 
      (a_{f_1} b_{f_2})^* (b_{f_1}a_{f_2}) \right)\right],
   \label{eq:Upsgen}
\end{eqnarray}
or equivalently,
\begin{eqnarray}
   \Gamma_{\Upsilon(4S)\to f_1f_2}(t_+,t_-) &=&  \label{eq:Upsgenpm} \\
     &&  \hspace{-3.5cm} {|c|^2\over4}e^{-\gamma_+ t_+}
\left[ e^{-\gamma_- t_-} |a_{f_1}b_{f_2}|^2 +
       e^{ \gamma_- t_-} |b_{f_1}a_{f_2}|^2 -
   2 \Re\left(
      (a_{f_1} b_{f_2})^* (b_{f_1}a_{f_2}) e^{i \delta m t_-}
      \right)\right], \nonumber
\end{eqnarray}
with
\begin{equation}
    t_\pm \equiv t_1 \pm t_2 \;,
\end{equation}
and we have used the relation $ 2 dt_1 dt_2 = dt_+ dt_- $.
Integrating (\ref{eq:Upsgen}) over $t_2$ and summing over all 
possible final states $f_2$, we obtain
\begin{equation}
 \begin{array}{l} {\displaystyle
  \Gamma_{\Upsilon(4S)\to f_1}(t_1) \equiv
  2 \sum_{f_2} \int_0^\infty \Gamma_{\Upsilon(4S)\to f_1f_2}(t_1,t_2)dt_2} \\
   {\displaystyle
  = |c|^2 \left[ |a_{f_1}|^2 e^{-\gamma_a t_1} +
                          |b_{f_1}|^2 e^{-\gamma_b t_1} -
     2\Re\left( e^{-(\gamma_+ - i \delta m) t_1} a_{f_1}^* b_{f_1}
      {\sum_{f_2}b_{f_2}^* a_{f_2}\over \gamma_+ + i \delta m}
         \right)\right] ,}
 \end{array}
\end{equation}
where we have used ({\ref{eq:ampnorm}), and the factor 2 arises from the fact
that the given final state can come from either side.
This together with (\ref{eq:BBdist}) and the Bell-Steinberger relation 
(\ref{eq:Bell-Stein}) establishes the general rule (\ref{eq:4Sinc}). 
The normalization (\ref{eq:Upsnorm}) and (\ref{eq:Ups1norm}) then
follows from (\ref{eq:4Sinc}) and (\ref{eq:BBnorm}).

Expressions for semileptonic decays are obtained by the substitutions
\begin{equation}
  \begin{array}{ll}
    a_{\ell^+} = p a_0\;,      &  b_{\ell^+} = p' a_0\; , \\
    a_{\ell^-} = q \bar a_0\;, &  b_{\ell^-} = -q' \bar a_0\; ,
  \end{array}
   \label{eq:lepamps}
\end{equation}
where we have used the assumption $\Delta B = \Delta Q$, and 
\begin{equation}
        a_0 \equiv Amp(B^0\to\ell^+)\;,\quad 
   \bar a_0 \equiv Amp(\bar B^0\to\ell^-)\; .
\end{equation}
Namely, (\ref{eq:BBdist}) gives
\begin{eqnarray}
   \Gamma_{B^0\to\ell^+}(t) &=& |c|^2 |a_0|^2 e^{-\gamma_+ t}
     \left[ |pq'|^2 e^{-\gamma_- t} + |p'q|^2 e^{\gamma_- t}
    + 2\Re\left((pq')^*(p'q) e^{i\delta m t}\right)\right],
                                     \nonumber \\
   \Gamma_{\ovl B^0\to\ell^-}(t) &=& |c|^2 |\bar a_0|^2 e^{-\gamma_+ t}
     \left[ |p'q|^2 e^{-\gamma_- t} + |pq'|^2 e^{\gamma_- t}
    + 2\Re\left((p'q)^*(pq') e^{i\delta m t}\right)\right],
                                     \nonumber \\
   \Gamma_{\ovl B^0\to\ell^+}(t) &=& 2|c|^2 |a_0|^2 |pp'|^2 e^{-\gamma_+ t}
     \left[ \cosh\gamma_- t - \cos \delta m t \right],
                                     \nonumber \\
   \Gamma_{B^0\to\ell^-}(t) &=& 2|c|^2 |\bar a_0|^2 |qq'|^2 e^{-\gamma_+ t}
     \left[ \cosh\gamma_- t - \cos \delta m t \right] . 
\end{eqnarray}
The $CPT$ relations (\ref{eq:CPTpq}) and (\ref{eq:CPTamp}) then lead to
(\ref{eq:BBlepdist}).

On $\Upsilon(4S)$, (\ref{eq:Upsgenpm}) and (\ref{eq:lepamps}) give the
di-lepton decay distributions:
\begin{eqnarray}
   \Gamma_{\Upsilon(4S)\to\ell^+\ell^-}(t_+,t_-) &=& \nonumber \\ 
   &&\hspace{-3cm} {|c|^2\over4} |\bar a_0 a_0|^2 e^{-\gamma_+ t_+}
     \left[ |pq'|^2 e^{-\gamma_- t_-} + |p'q|^2 e^{\gamma_- t_-}
    + 2\Re\left((pq')^*(p'q) e^{i\delta m t_-}\right)\right],
                                     \nonumber \\
   \Gamma_{\Upsilon(4S)\to\ell^-\ell^+}(t_+,t_-) &=& \nonumber \\
   &&\hspace{-3cm}  {|c|^2\over4} |\bar a_0 a_0|^2 e^{-\gamma_+ t_+}
     \left[ |p'q|^2 e^{-\gamma_- t_-} + |pq'|^2 e^{\gamma_- t_-}
    + 2\Re\left((p'q)^*(pq') e^{i\delta m t_-}\right)\right],
                                     \nonumber \\
   \Gamma_{\Upsilon(4S)\to\ell^+\ell^+}(t_+,t_-) &=& 
     {|c|^2\over2} |a_0|^4 |pp'|^2 e^{-\gamma_+ t_+}
     \left[ \cosh\gamma_- t_- - \cos \delta m t_- \right],
                                     \nonumber \\
   \Gamma_{\Upsilon(4S)\to\ell^-\ell^-}(t_+,t_-) &=& 
     {|c|^2\over2} |\bar a_0|^4 |qq'|^2 e^{-\gamma_+ t_+}
     \left[ \cosh\gamma_- t_- - \cos \delta m t_- \right].
\end{eqnarray}
Note that the opposite-sign lepton rates satisfy the relation
\begin{equation}
     \Gamma_{\Upsilon(4S)\to\ell^+\ell^-}(t_+,t_-) =
     \Gamma_{\Upsilon(4S)\to\ell^-\ell^+}(t_+,-t_-) \;,
\end{equation}
which corresponds to re-labeling the final states.
Under $CPT$ symmetry this simplifies to
\begin{eqnarray}
    \Gamma_{\Upsilon(4S)\to\ell^+\ell^-}(t_+,t_-) &=& 
    \Gamma_{\Upsilon(4S)\to\ell^-\ell^+}(t_+,t_-) \nonumber  \\
     &=& {A_0^4\over8}  e^{-\gamma_+ t_+}
       \left[ \cosh\gamma_- t_- + \cos \delta m t_- \right], 
                                               \nonumber     \\
    \Gamma_{\Upsilon(4S)\to\ell^+\ell^+}(t_+,t_-) &=& 
         {|p|^2\over|q|^2}{A_0^4\over8}  e^{-\gamma_+ t_+}
       \left[ \cosh\gamma_- t_- - \cos \delta m t_- \right],
                                                             \\ 
    \Gamma_{\Upsilon(4S)\to\ell^-\ell^-}(t_+,t_-) &=& 
         {|q|^2\over|p|^2}{A_0^4\over8}  e^{-\gamma_+ t_+}
       \left[ \cosh\gamma_- t_- - \cos \delta m t_- \right] .
                                               \nonumber
\end{eqnarray}
Integrating over $t_+$ from $|t_-|$ to $\infty$ gives (\ref{eq:Upstmi}).
Note that $CPT$ invariance ensures that the decay distributions are
symmetric under sign change of $t_-$. Thus, such asymmetry (for example,
if the $\ell^+$ side tends to decay earlier then the $\ell^-$ side
in the $\ell^+\ell^-$ sample) is a
signature of $CPT$ violation \cite{CPT-B}. 
This is in contrast to the case of
lepton tagged $CP$ eigenstates (e.g., $\Psi K_s$ \cite{PsiKs,NirQuinn}) 
where the asymmetry
with respect to $t_-$ is possible under $CPT$ invariance and 
signals $CP$ violation.


\begin{thebibliography}{99}
\bibitem{CPTamp}
  Under $CPT$ symmetry, the $Amp(B^0\to l^+X)$ can be taken as the 
  complex conjugate of $Amp(\ovl B^0\to l^-\ovl X)$ where all the
  helicities are flipped while keeping the momenta unchanged, 
  but only to the first order
  in the perturbation theory. However, after summing over helicities,
  and over the possible state $X$, the relation 
  effectively holds to a high accuracy.
  Such incoherent sum (as well as the sum over electron and
  muon modes) is allowed since the square of the amplitude
  factors out in all the relevant decay time distributions.
\end{thebibliography}
\end{document}